%% file: main.tex
\documentclass{llncs}

\usepackage[T1]{fontenc}
\usepackage{graphicx}
\usepackage{hyperref}
\usepackage{color}
\usepackage[style=lncs]{biblatex}
\usepackage{duckuments}
\usepackage{siunitx}
\usepackage{todonotes}
\usepackage[nolist]{acronym}
\usepackage{listings}
\usepackage{subcaption}
\usepackage{booktabs}
\usepackage{multirow}
\usepackage[table,x11names]{xcolor}
\usepackage{tabularx}
\usepackage{pgfplots}

\graphicspath{{assets/images/}}

\begin{document}

\title{Towards Delta Aware Training: Efficient DNN Weight Storage for Resource-Constrained FPGAs}
\titlerunning{Towards Delta Aware Training}

\author{
    David Peter Federl\inst{1}\orcidID{0009-0000-8542-7457} \and \\
    Lukas Einhaus\inst{1}\orcidID{0000-0002-6102-7077} \and \\
	Andreas Erbslöh\inst{1}\orcidID{0000-0001-6702-892X} \and \\
	Gregor Schiele\inst{1}\orcidID{0000-0003-4266-4828}
}
\authorrunning{DP Federl, et. al.}
\institute{
	Intelligent Embedded Systems Lab, University Duisburg-Essen, Duisburg, Germany \\
	\email{\{david-peter.federl,lukas.einhaus,andreas.erbsloeh,gregor.schiele\}@uni-due.de} \\
	\url{https://uni-due.de/es/en/en_home.php}
}

\maketitle

\begin{acronym}
	\acro{ml}[ML]{machine learning}
    \acro{mac}[MAC]{multiply-and-accumulate}
	\acro{dnn}[DNN]{deep neural network}
	\acro{fpga}[FPGA]{field programmable gate array}
	\acro{gpu}[GPU]{graphics processing unit}
	\acro{tpu}[TPU]{tensor processing unit}
	\acro{qat}[QAT]{quantization-aware training}
	\acro{mlp}[MLP]{multi-layer perceptron}
	\acro{sota}[SOTA]{State-of-the-Art}
	\acro{msb}[MSB]{most significant bit}
	\acro{lsb}[LSB]{least significant bit}
	\acro{htanh}[hard TanH]{hard hyperbolic tangent}
	\acro{tanh}[TanH]{hyperbolic tangent}
	\acro{bram}[BRAM]{block random access memory}
	\acro{dat}[DAT]{delta-aware training}
\end{acronym}

\input{sections/0_Abstract}
\acresetall
\input{sections/1_Introduction}
\acresetall
\input{sections/2_Related_Work}
\input{sections/3_Delta_Compression}
\input{sections/4_Evaluation}
\input{sections/5_Hardware_Utilization}
\input{sections/6_Discussion}
\input{sections/7_Acknowldigements}
\printbibliography

\end{document}

%% file: sections/0_Abstract.tex
\begin{abstract}
    The deployment of embedded deep neural networks on resource-constrained \acp{fpga} is challenging due to limited memory and computational capacities. 
    We introduce a new compression technique to reduce the memory footprint by saving weights in deltas with lower bitwidth and training the network to cope with compressed deltas.
    Two delta schemes are investigated: consecutive deltas and deltas with a fixed-reference value.
    We evaluate both on the FashionMNIST data set with a multi-layer-perceptron. 
    The results indicate that fixed-reference delta compression outperforms the consecutive variant, achieving a validation accuracy of approximately \SI{78.6}{\percent}, with \SI{4}{\bit} weight deltas, representing an accuracy loss of roughly \SI{8.3}{\percent} compared to a fixed-point network with \SI{8}{bit}. 
    Our specialized hardware accelerator with a delta-compressed multiply-and-accumulate operator compresses weights by nearly \SI{50}{\percent} and achieves a maximum throughput of \num{7.992}M~MACs/s on an AMD Spartan-7 S15 \ac{fpga}.
    
	\keywords{embedded deep neural networks, resource-constrained, field programmable gate array, hardware accelerator, fixpoint-arithmetic, quantization aware training, weight compression}
\end{abstract}

%% file: sections/1_Introduction.tex

\section{Introduction}

Deploying \acp{dnn} on embedded hardware with restricted resources is becoming more and more common, e.g., due to privacy concerns or hard realtime requirements. 
Model compression techniques in combination with special hardware can help to run them fast and efficiently.
An example for such special hardware are low-power \acp{fpga}.
The benefit of \acp{fpga} is that they can be programmed as application specific hardware accelerators for neural network computations instead of being restricted to general purpose computation acceleration as provided by standard hardware such as \acp{gpu}.
However, the full accelerator is required to fit the \ac{fpga} itself to provide reasonable inference speeds in embedded systems.

In order to extend the capacity limitations of such systems we propose a delta based compression technique that allows us to decrease the needed memory by storing the weights of the neural network with reduced bits. During inference, the deltas are reconstructed in the desired resolution.
We evaluate our performance on the FashionMNIST data set with a \ac{mlp}.
Our concrete contributions are as follows: \begin{itemize}
	\item We implement a \ac{dat} approach for fixed-point \acp{dnn}.
    \item We evaluate the performance and implementation of two delta schemes (consecutive and fixed-reference) on FashionMNIST. 
    \item We contribute a PyTorch-suitable implementation for delta compressed Linear Layers with batch-normalized input.
	\item We evaluate the proposed hardware accelerator with regards to \ac{fpga} utilization, inference latency, and loss of accuracy of the neural network compared to full-precision and pure fixed-point quantized implementations.
\end{itemize}

The paper is structured as follows: We discuss related work in~Section~\ref{sec:related-work}, before presenting our approach for~\ac{dat} in~Section~\ref{sec:delta-compression}.
We evaluate our approach in~Section~\ref{sec:evaluation} and discuss the limitations. Section~\ref{sec:hardware} shows the FPGA implementation and hardware costs for the delta-based MAC operator. Section~\ref{sec:conclusion} concludes the paper and proposes future work. 

%% file: sections/2_Related_Work.tex
\section{Related Work}\label{sec:related-work}

Related work can be grouped into two categories.
The first group contains work related to the general application of delta-based compression techniques mainly focusing on file based storage systems such as backups, version control systems, or image compression.
The second group focuses on delta-based compression techniques as a tool to reduce the storage size of training snapshots \acp{dnn}.

\subsection{Delta Compression Algorithms}

Delta compression techniques~\cite{sayood_lossless_2003} typically operate at the level of strings or bytes, encoding differences between them.
The most well-known application for delta-based compression is the JPEG image format.
To further improve the performance of such compression algorithms they are often supplemented with additional means of compression like dictionary based data encoding, as demonstrated by Lempel-Ziv-Markov chains.

Another well known application for delta based compression at the file level are \ac{sota} compression tools in operating systems and version control systems~\cite{hunt_delta_1998}.
Incremental backup systems also deploy delta based archives, so that only updates to the first complete backup are stored.

\subsection{Deep Neural Network Compression}

Due to the fact that the size of \acp{dnn} is continuously increasing in recent years, there are also many attempts to decrease the amount of storage consumed for these neural networks, especially during training.

Some of these approaches utilize quantization to reduce the amount of storage required for each weight.
Quantization, however, always causes an inevitable information loss for the neural network, since the discrete steps between different values are much larger than the resolution of 32-bit floating-point values.
Therefore, quantization is often combined with \ac{qat}, which emulates the quantization of the target data-type in the forward step during the training process, whereas the backward step uses standard floating-point values, so that gradients can be calculated effortlessly.
The benefit of quantization is the fact that the network can be trained with weights targeting a specific hardware platform, which can yield a big performance boost on platforms, like embedded systems without native floating-point support.

During neural network training, it is often necessary to store snapshots of each weight to later dissect the evolution during the training process to understand the updates during training.
To minimize the amount of storage required for these snapshots, multiple compression techniques were evaluated.
\citetitle{hu_deltadnn_2020}~\cite{hu_deltadnn_2020} calculates the deltas between different versions of a \ac{dnn} and compresses the delta with a lossy error bound compression that is limited by the lowest acceptable loss during inference.
\citetitle{zhang_qdcompressor_2021}~\cite{zhang_qdcompressor_2021} enhances this by taking local differences in floating-point ranges inside a layer into account and applying the quantization before the delta computation~\cite{jin_design_2023}.

The neural network weight base-delta-immediate~(NNW-BDI) compression scheme~\cite{bersatti_neural_2021} optimizes the BDI compression proposed by~\cite{pekhimenko_practical_2016} for \acp{dnn} by adapting it for usage with highly dynamic floating-point numbers.
BDI itself uses data similarity between values that are close to each other in memory. 
NNW-BDI can achieve a compression of~\SI{85}{\percent} without significant accuracy loss.

This paper presents a delta-based compression of neural network weights to decrease memory for resource-constrained embedded hardware.
The related work, however, focuses on compressing snapshots of floating-point models, whereas we are focusing on compressing deployment weights of fixed-point models.
Initially, we investigate methods for a quantization-aware and delta-aware training using \ac{mlp} on FashionMNIST.

%% file: sections/3_Delta_Compression.tex
\section{Weight Delta Compression for DNNs}\label{sec:delta-compression}

Our approach for delta compression of weights is a two-step process that involves a series of operations aimed at reducing the memory footprint of the weights while preserving their information.
Note that the order in which these two steps are performed does matter for the result.
We choose to keep the order as follows for the remainder of this work: \begin{enumerate}
	\item First, the \textbf{delta calculation}
	\item Then, the \textbf{data compression}
\end{enumerate}
Each of these steps can then be performed in a multitude of different variations.
The following sections explain the different options used.

\subsection{Delta Calculation}

We use two different types of delta calculation: a) consecutive and b) fixed-reference delta calculation.
Both types are computed per-layer.
To facilitate the calculation of deltas, it is essential to establish a fixed order for the weights.
We flattened the tensor with the weights shown in~\autoref{lst:delta-consecutive} and~\autoref{lst:delta-fixed} to establish this fixed order.
Given that the shape of the tensor is crucial for its subsequent use as weights within the neural network, it must be preserved.
The solution we used for our implementations is to restore the original tensor shape after the deltas are computed.

\subsubsection{Consecutive.}

\begin{figure}[!ht]
	\centering
    \includegraphics[width=0.95\textwidth]{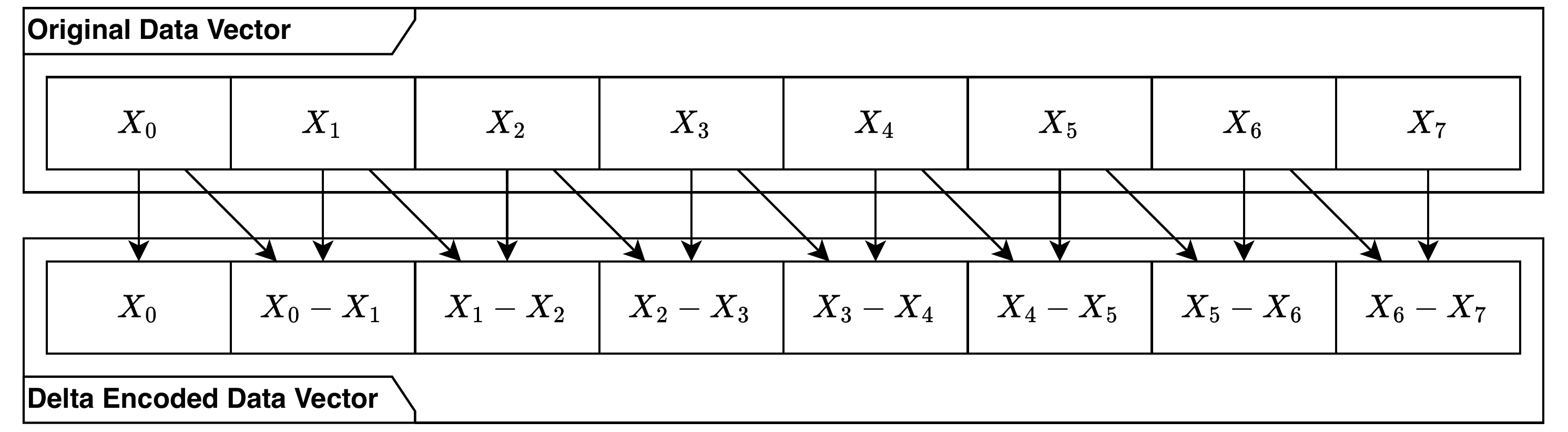}
    \caption{Consecutive Delta}\label{fig:consecutive-delta}
\end{figure}

\begin{lstlisting}[
  language=Python, 
  firstline=4,
  caption={Consecutive Delta Calculation},
  captionpos=b,
  label={lst:delta-consecutive},
  float=!ht
]
def delta(self, input: Tensor) -> Tensor:
    original_shape = input.shape
    input = input.flatten()
    for index in reversed(range(1, len(input))):
        input[index] = input[index] - input[index - 1]
    input = input.reshape(original_shape)
    return input
\end{lstlisting}

Consecutive deltas are computed by determining the delta between two adjacent elements within the weight tensor of the selected layer.
Given that the calculation of a delta requires at least two values, the first value of the tensor remains unchanged by the delta computation, it will be called the~\textit{reference value} in the following sections.
\autoref{fig:consecutive-delta} shows a visualization of this schema.

\subsubsection{Fixed.}

\begin{figure}[!ht]
    \centering
    \includegraphics[width=\textwidth]{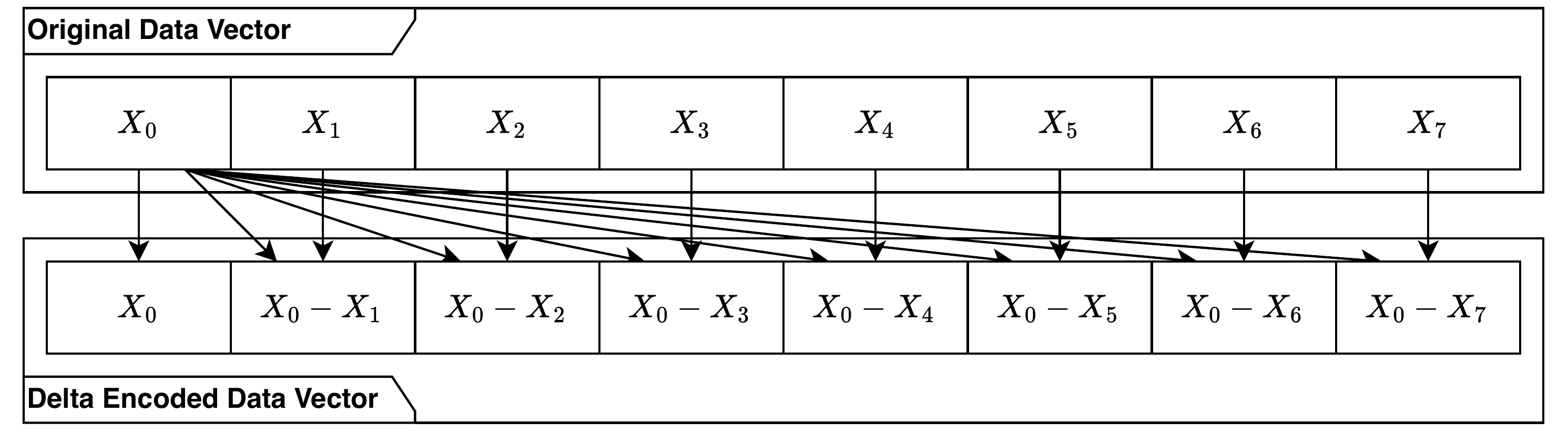}
    \caption{Fixed Reference Delta}\label{fig:fixed-delta}
\end{figure}


\begin{lstlisting}[
  language=Python, 
  firstline=4,
  caption={Fixed Delta Calculation},
  captionpos=b,
  label={lst:delta-fixed},
  float=!ht
]
def delta(self, input: Tensor) -> Tensor:
    original_shape = input.shape
    input = input.flatten()
    input[1:] -= input[0]
    input = input.reshape(original_shape)
    return input
\end{lstlisting}

Fixed reference deltas are computed by determining all deltas within a layer, using a fixed-reference value.
The benefit of this approach is the fact that the errors are not propagated through the weight tensor, since all delta calculations are independent of each other.
The fixed reference remains unchanged by the delta computation.
We refer to it as \textit{reference value} in the remainder of this paper. 
This approach is visualized in~\autoref{fig:fixed-delta}.

\subsection{Data Compression}

After computing~$n$-bit delta values, we compress all deltas to a given (smaller) bitwidth~$m$, for example, encoding our delta in $m=4$ bit instead of $n=8$ bit. 

\begin{figure}[!ht]
	\centering
	\includegraphics[width=0.75\textwidth]{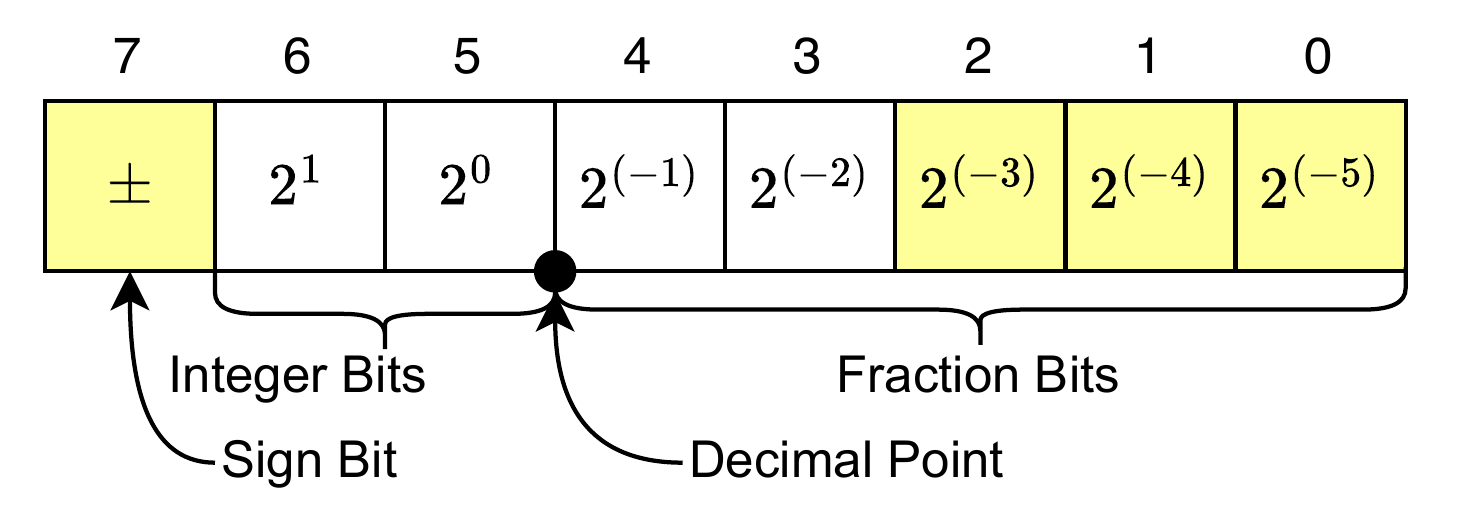}
	\caption{Fixed-Point Representation}\label{fig:fxp-representation}
\end{figure}

To allow positive and negative deltas, we always include the sign bit. 
For the remaining $m-1$ bits, we distinguish between two cases.
In the first case, the full delta value can be encoded in the $m-1$ least significant bits of the uncompressed delta.
In this case, we directly include these bits in our compressed delta. 
Thus, for 4-bit compressed values from 8-bit deltas, we would choose bits 0, 1, 2, and 7 (see \autoref{fig:fxp-representation}). 
In the second case, the uncompressed delta is too large to fit into~$m-1$ bit. 
In this case, we use saturation semantics, for example, the maximum/minimum $m-1$ bit value. 
In our example, we would choose~$0111$ for positive deltas and~$1001$ for negative deltas (two's complement). 
When applying deltas, they are first expanded to signed $n$ bit numbers and then added to the reference value. 

We experimented with other variants, but found that they did not perform as well. 
First, we directly took the selected bits without saturation semantics. 
The resulting networks often did not train at all. 
Second, we added a selectable offset, allowing one to shift the selected bits away from bit 0. 
As an example, with an offset of 2, we would select bits 2, 3, 4. 
However, we did not find offsets that performed better. 
Therefore, we also abandoned this idea.

%% file: sections/4_Evaluation.tex
\section{Experimental Evaluation}\label{sec:evaluation}

This section describes the experimental setup, the used dataset, the used \ac{mlp} model and measurements for different settings.

\subsection{Network Architecture and Training Methodology}

We use the FashionMNIST dataset~\cite{xiao_fashionmnist_2017}.
It consists of~\num{70000} grayscale images with~28 by~28 pixels.
Each image contains values in the range of~zero to~255 and shows a variety of different clothing pieces that are grouped into ten different categories.
It is split into~\num{60000} training images and~\num{10000} test images, which are sampled to represent the actual distribution of samples of the data sets.

\begin{figure}[!ht]
	\centering
	\includegraphics[width=\textwidth]{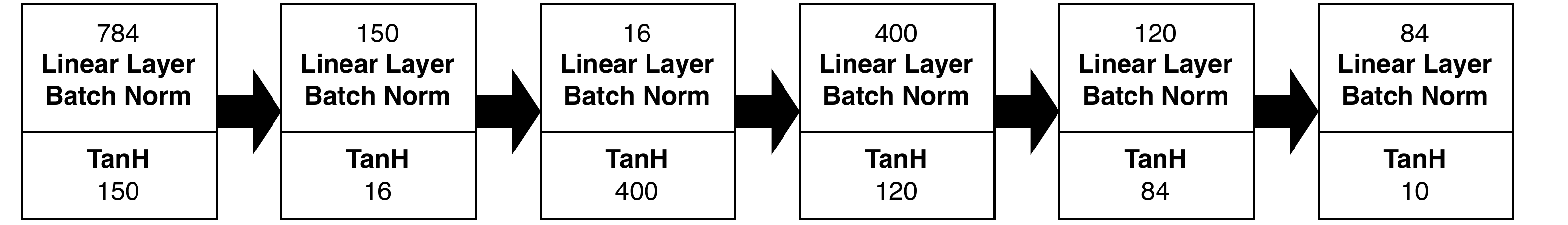}
	\caption{Structure of the proposed Neural Network}
    \label{fig:network}
\end{figure}

The images are preprocessed by min-max scaling to have all data in the range of $[-1.0, +1.0]$ and flattened to be used with the network shown in~\autoref{fig:network}.

The neural network as shown in~\autoref{fig:network} consists of six layers with~\num{185320} weights in total.
Each linear layer is followed by a batch norm and a \ac{htanh} activation.
We choose \ac{htanh} instead of \ac{tanh} as its lower complexity leads to simpler and faster hardware implementations.

The realization and training of the neural networks is based on the open source elasticAI ecosystem~\cite{einhaus_insitu_2021}.
The training uses PyTorch-based models in combination with fixed-point layers provided by the~\textit{elasticAI.creator}\footnote{\url{https://github.com/es-ude/elastic-ai.creator}}, our tool for the training and generation of hardware-accelerated \acp{dnn}.
We further extend the \textit{elasticAI.creator} with a \ac{dat} implementation and a specialized hardware template for delta-based MAC operators to generate HW accelerators.

We train all networks on a MacBook Pro with an M4 Max Chip and~\SI{128}{\giga\byte} unified RAM, using only its standard cores.
Throughout our experiments, we use Adam as optimizer and choose CrossEntropyLoss as loss function.
We train for~100 epochs with a batch size of~512 and a learning rate of~\num{0.001}.
We repeat the training process for each configuration~100~times and report the mean metrics across all runs.

\subsection{Baseline Performance}

As a general baseline we first train a full-resolution (\SI{32}{\bit} floating-point precision) model.
Its mean validation accuracy is approximately~\SI{87}{\percent}.
We then train a number of fixed-point networks. 
To denote the fixed-point type we use the notation Q$n$.$m$, where $n$ represents the number of integer bits and $m$ is the number of fraction bits.
Since we add one additional bit for the sign, the total bits required for each fixed-point number can be calculated by $n+m+1$. 
We train each possible \SI{8}{\bit} fixed-point quantization configuration (Q0.7, Q1.6, Q2.5, Q3.4, Q4.3, Q5.2 and Q6.1).
The results are shown in \autoref{tab:baseline}.
After training for~100~epochs the accuracy of the three best configurations (Q0.7, Q1.6, Q2.5) is saturated, while we can detect that other networks (Q3.4, Q4.3, Q5.2, Q6.1) do not learn, since the loss and accuracy stay the same for many epochs.
This is likely due to the insufficient number of fractional bits.
The accuracy of two of the configurations (Q1.6, Q2.5) is comparable to the floating-point network.

\begin{table}[!ht]
	\centering
	\caption{Average accuracy over 100 training runs with 100 epochs}\label{tab:baseline}
	\vskip\baselineskip
    \addtolength{\tabcolsep}{5pt} 
	\begin{tabularx}{0.9\textwidth}{ >{\raggedleft\arraybackslash}X | r r | r r }
		\toprule
		\multirow[b]{2}{*}{\textbf{Parameter Type}}          & \multicolumn{2}{c|}{\textbf{Validation}} & \multicolumn{2}{c}{\textbf{Training}}                                       \\
		                                                     & \textbf{Accuracy}                       & \textbf{Loss}                         & \textbf{Accuracy}   & \textbf{Loss} \\
		\midrule
		\rowcolor{lightgray} Floating-Point \SI{32}{\bit}    & \SI{87}{\percent}                       & \num{0.7716}                          & \SI{98.9}{\percent} & \num{0.0293}  \\
		Fixed-Point \SI{8}{\bit} (Q0.7)                      & \SI{86}{\percent}                       & \num{1.0074}                          & \SI{89}{\percent}   & \num{0.9208}  \\
		\rowcolor{lightgray} Fixed-Point \SI{8}{\bit} (Q1.6)                      & \SI{86}{\percent}                       & \num{0.5930}                          & \SI{98}{\percent}   & \num{0.2084}  \\
		\rowcolor{lightgray} Fixed-Point \SI{8}{\bit} (Q2.5) & \SI{87}{\percent}                       & \num{0.5902}                          & \SI{99}{\percent}   & \num{0.489}   \\
		Fixed-Point \SI{8}{\bit} (Q3.4)                      & \SI{10}{\percent}                       & \num{5.2691}                          & \SI{10}{\percent}   & \num{2.3028}  \\
		Fixed-Point \SI{8}{\bit} (Q4.3)                      & \SI{10}{\percent}                       & \num{2.3026}                          & \SI{10}{\percent}   & \num{2.3027}  \\
		Fixed-Point \SI{8}{\bit} (Q5.2)                      & \SI{10}{\percent}                       & \num{2.3026}                          & \SI{10}{\percent}   & \num{2.3026}  \\
		Fixed-Point \SI{8}{\bit} (Q6.1)                      & \SI{10}{\percent}                       & \num{2.3026}                          & \SI{10}{\percent}   & \num{2.3026}  \\
		\bottomrule
	\end{tabularx}
\end{table}

To provide a better comparison with the 4-bit delta network, we also trained neural networks with 4-bit fixed-point numbers, but they did not converge. 
We therefore dropped this approach and limit ourselves to comparing our delta compression to eight-bit fixed-point networks.

\subsection{Post-Training Delta}

Initially, we applied our delta compression algorithms to pre-trained fixed-point networks, i.e., using a post training approach. 
However, this yielded unsatisfactory results, resulting in the degradation of the networks to a point where all trained information was lost.
The accuracy of the model decreased to approximately~\SI{10}{\percent} and the loss increased to~\num{2.3}, which is equivalent to the loss of a random guess.
 To overcome this, we introduced \acl{dat}.

\subsection{Delta-Aware Training}

Our \acl{dat} was implemented with the~\textit{elasticAI.creator} framework.
The fixed-point arithmetic of the framework is used to incorporate delta compression, as elaborated in Section~\ref{sec:delta-compression}.
The following sections present measurement results for our two different delta compression schemas: i) consecutive and ii) fixed-reference delta compression.

\begin{table}[!ht]
    \centering
    \caption{Comparison of validation accuracy and storage requirements for different network types}\label{tab:accuracy-comparison}
    \addtolength{\tabcolsep}{5pt} 
    \begin{tabular}{lcccc}
        \toprule
        & \textbf{32-bit} &\textbf{Q2.5, 8-bit}       & \textbf{Q2.5, 4-bit}  & \textbf{Q2.5, 4-bit}  \\
        & &w/o delta-compr.             & fixed-reference       & consecutive           \\ \midrule
         Accuracy & \SI{87}{\percent} &\SI{87}{\percent}  & \SI{78.66}{\percent}  & \SI{75.99}{\percent}  \\
        Weight Size & 741.2 KB & 185.3 KB & 94.9 KB & 94.9 KB \\\bottomrule
    \end{tabular}
\end{table}

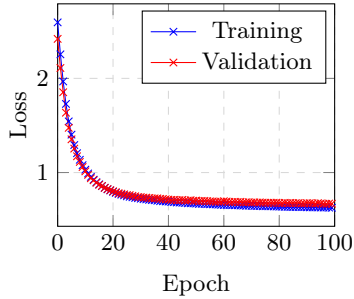
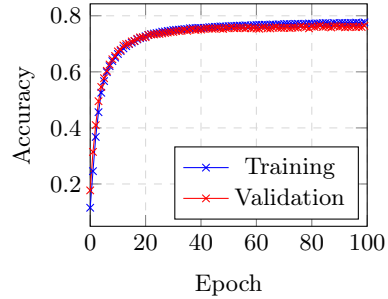
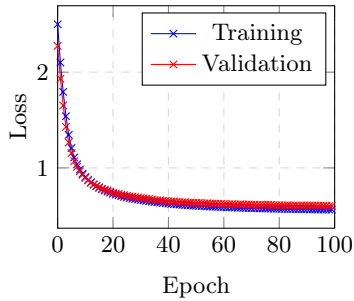
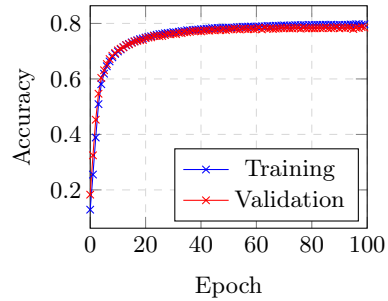
\begin{figure}[!ht]
	\centering
	\begin{subfigure}{0.43\textwidth}
        \begin{tikzpicture}
            \begin{axis}[
              width=\linewidth, 
              grid=major,
              grid style={dashed,gray!30}, 
              xlabel=Epoch, 
              xmax=100,
              xmin=0,
              ylabel=Loss,
              ylabel style = {yshift=-20pt},
              legend pos=north east
            ]
            \addplot[mark=x, color=blue] table[x=epoch,y=training,col sep=comma] {assets/data/q2.5_d4o0_consecutive_bs512_loss.csv}; 
            \addlegendentry{Training}
            
            \addplot[mark=x, color=red] table[x=epoch,y=validation,col sep=comma] {assets/data/q2.5_d4o0_consecutive_bs512_loss.csv}; 
            \addlegendentry{Validation}
          \end{axis}
      \end{tikzpicture}
		\caption{Loss (Consecutive Delta)}\label{fig:dc-consecutive-loss}
	\end{subfigure}
	\hfill
	\begin{subfigure}{0.43\textwidth}
        \begin{tikzpicture}
            \begin{axis}[
              width=\linewidth, 
              grid=major,
              grid style={dashed,gray!30}, 
              xlabel=Epoch, 
              xmax=100,
              xmin=0,
              ylabel=Accuracy,
              ylabel style = {yshift=-10pt},
              legend pos=south east
            ]
            \addplot[mark=x, color=blue] table[x=epoch,y=training,col sep=comma] {assets/data/q2.5_d4o0_consecutive_bs512_accuracy.csv}; 
            \addlegendentry{Training}
            
            \addplot[mark=x, color=red] table[x=epoch,y=validation,col sep=comma] {assets/data/q2.5_d4o0_consecutive_bs512_accuracy.csv}; 
            \addlegendentry{Validation}
          \end{axis}
      \end{tikzpicture}
		\caption{Accuracy (Consecutive Delta)}\label{fig:dc-consecutive-accuracy}
	\end{subfigure}
	\vskip\baselineskip
	\begin{subfigure}{0.43\textwidth}
        \begin{tikzpicture}
            \begin{axis}[
              width=\linewidth, 
              grid=major,
              grid style={dashed,gray!30}, 
              xlabel=Epoch, 
              xmax=100,
              xmin=0,
              ylabel=Loss,
              ylabel style = {yshift=-20pt},
              legend pos=north east
            ]
            \addplot[mark=x, color=blue] table[x=epoch,y=training,col sep=comma] {assets/data/q2.5_d4o0_fixed_bs512_loss.csv}; 
            \addlegendentry{Training}
            
            \addplot[mark=x, color=red] table[x=epoch,y=validation,col sep=comma] {assets/data/q2.5_d4o0_fixed_bs512_loss.csv}; 
            \addlegendentry{Validation}
          \end{axis}
        \end{tikzpicture}
		\caption{Loss (Fixed Reference Delta)}\label{fig:dc-fixed-loss}
	\end{subfigure}
	\hfill
	\begin{subfigure}{0.43\textwidth}
        \begin{tikzpicture}
            \begin{axis}[
              width=\linewidth, 
              grid=major,
              grid style={dashed,gray!30}, 
              xlabel=Epoch, 
              xmax=100,
              xmin=0,
              ylabel=Accuracy,
              ylabel style = {yshift=-10pt},
              legend pos=south east
            ]
            \addplot[mark=x, color=blue] table[x=epoch,y=training,col sep=comma] {assets/data/q2.5_d4o0_fixed_bs512_accuracy.csv}; 
            \addlegendentry{Training}
            
            \addplot[mark=x, color=red] table[x=epoch,y=validation,col sep=comma] {assets/data/q2.5_d4o0_fixed_bs512_accuracy.csv}; 
            \addlegendentry{Validation}
          \end{axis}
        \end{tikzpicture}
		\caption{Accuracy (Fixed Reference Delta)}\label{fig:dc-fixed-accuracy}
	\end{subfigure}
	\caption{Average Neural Network Metrics for our delta-aware training (\ac{dat}) on the FashionMNIST dataset for Q2.5}\label{fig:dc-metrics}
\end{figure}

\subsubsection{Consecutive Delta Compression.}

The consecutive delta compression enables the network to use weights that are further away from the reference value than a single delta calculation would allow.
However, this is only important if the deltas slowly move outside of the compression range, which we did not observe for the network used for our evaluation.

In this particular case, the network has been demonstrated to converge to a valid solution, as illustrated in \autoref{fig:dc-consecutive-loss} and  \autoref{fig:dc-consecutive-accuracy}.
Nevertheless, the network is incapable of attaining an equivalent performance to that of fixed reference delta compression.
In addition to slightly lower accuracy compared to the implementation of a fixed-reference, the execution speed during training is \num{27} times higher, with \SI{81}{\second} per epoch.

\subsubsection{Fixed Reference Delta Compression.}

Fixed delta compression enables the network to utilize weights that approximate the reference value.
This behavior is advantageous for the network, as the weights are close to the reference value.
The network converges to a valid solution, as illustrated in \autoref{fig:dc-fixed-loss} and \autoref{fig:dc-fixed-accuracy}.
However, it is important to acknowledge that the network cannot achieve the same level of accuracy as the eight-bit fixed-point network.
This results in a decrease in accuracy of approximately \SI{8.3}{\percent} to \SI{78.6}{\percent}.
The execution speed during training is nearly as fast as the fixed-point network or floating-point network, with \SI{3}{\second} per epoch.

%% file: sections/5_Hardware_Utilization.tex
\section{Hardware Utilization}\label{sec:hardware}

We analyze the hardware utilization of the MAC operator implementation used in the \ac{mlp} accelerator.
We report the timing characteristics and resource utilization as estimated by AMD Vivado 2024.2.
We deploy the resulting accelerators on the~\textit{elasticAI.hardware}~(ENV5)~\cite{schiele_elastic_2019}, which features a microcontroller and the AMD Spartan-7 S15~(xc7s15ftgb196-2) \ac{fpga}. 
The Verilog implementation is generated using the \verb|mac| plugin of the~\textit{elasticAI.creator}. 
All designs include a SPI interface and a middleware to interact with the accelerators from the microcontroller.

\begin{table}[ht] \centering
    \caption{Overview of the hardware costs and performance from different MAC implementations with different number of parallel-used multipliers at common number of weights of~84, 4-bit for delta compression and 8-bit for data processing.}
    \vskip\baselineskip
    \addtolength{\tabcolsep}{2pt} 
    \begin{tabular}{lccccccc}
    \toprule \textbf{MAC type} & \textbf{\#LUTs} & \textbf{\#FFs}  & \textbf{\#FFs}   & \textbf{\#DSPs}  & \textbf{\#Cycles}  & \textbf{$f_{max}$}  & \textbf{MACs/s}        \\
                      &          & (buffer)         & (params)  &           &             &{[}MHz{]}  &  \\ \midrule
    normal            & 489     & 39        & 1352      & 1         & 86          & 186.57    & 2.169M       \\
    consecutive       & 412     & 47        & 1024      & 1         & 86          & 177.30    & 2.062M      \\
    fixed-reference   & 401     & 36        & 1024      & 1         & 86          & 182.48    & 2.122M      \\ \midrule
    normal            & 924     & 47        & 1352      & 2         & 44          & 185.18    & 4.209M      \\
    consecutive       & 726     & 43        & 1024      & 2         & 44          & 181.16    & 4.117M      \\
    fixed-reference   & 702     & 41        & 1024      & 2         & 44          & 182.48    & 4.147M      \\ \midrule
    normal            & 986     & 35        & 1352      & 4         & 23          & 185.18    & 8.051M      \\
    consecutive       & 834     & 39        & 1024      & 4         & 23          & 179.86    & 7.819M      \\
    fixed-reference   & 799     & 39        & 1024      & 4         & 23          & 183.82    & 7.992M      \\ \midrule
    \end{tabular}
    \label{tab:util}
\end{table}

Table~\ref{tab:util} shows the hardware costs for (i) uncompressed / normal, (ii) delta-based using consecutive reconstructed and (iii) delta-based using fixed-referenced MAC accelerators with different numbers of multipliers.
Optionally, the DSP slices can be replaced with custom Verilog multipliers.
All MAC designs are buffering the data to the DSP slices to enable high-throughput with parametrized number of parallel-used multipliers.
The reconstruction of the weights takes place during the pipelining process. Here, each delta value has to be expanded via two's complement, starting with LSB.

\begin{equation}\label{eq:compression}
    \text{compression rate (CR)} = 1-\frac{\overbrace{\text{bitwidth}}^{\text{initial}} + \overbrace{\text{deltawidth} * \text{\#params}}^{\text{compressed weights}}}{\underbrace{\text{bitwidth} * \text{\#params}}_{\text{original weights}}}
\end{equation}

Compared to the normal MAC implementation, the delta-based consecutive and fixed-reference implementation use less resources.
With int8 processing and a delta width of 4-bit, the weights are compressed by~\SI{48.81}{\%} using~(\ref{eq:compression}).
Each implementation needs $\text{ceil}(\#params / \#mult) + 2~cycles$.
The maximum clock frequency is extracted using the worst negative slack which is typical for the used AMD FPGA. 
The resource utilization of the MAC implementations scales linear with the number of weights and the number of FFs for weight storing scales with the compression rate in~(\ref{eq:compression}).
In general, the delta-compressed MAC operator using the fixed-reference method is showing a lower utilization due to the easier reconstruction calculation.

Applying the delta-based MAC operator in neural networks allows to increase the throughput using a single-port \ac{bram} for weight storing.
If a delta bitwidth of 4-bit is applied on 8-bit data processing or 8-bit delta on 12-bit input, the throughput is doubled due to getting two values or one value in each 8-bit cell read-out. 
It is important to note that LUT resource consumption increases significantly when more than one multiplier is used for parallel computation.
This is due to the unrolling of the data slicing and the parallel buffers and adders.

%% file: sections/6_Discussion.tex
\section{Conclusion and Future Work}\label{sec:conclusion}

In this paper, we study the effectiveness of using delta compression to reduce the memory required to store the weights of a neural network on an \ac{fpga}. This allows to deploy bigger \acp{dnn}. 
Of the two studied variants, fixed reference delta compression outperforms consecutive delta compression with less accuracy loss and a more resource-efficient hardware implementation. Clearly, since we only use a single data set with a single neural network, our results just provide an indication that the concept is applicable.

The proposed algorithm provides just the first step into delta-based compression of neural networks for low-resource \acp{fpga}.
In future work we plan to evaluate further bit selection schema, such as bucket based compression, and stochastic rounding based value computation which should provide better solutions for low bit widths, as indicated by~\cite{buron_reducing_2025}. 

In order to improve the performance of the fixed-reference delta compression and lead the training to a faster convergence, an optimizer employing a weight decay can be used to move the weights altogether closer to zero.
In addition to these improvements on the practical site, more neural network models with different datasets should be evaluated in order to gain a greater understanding of the possible limitations for delta compression.

%% file: sections/7_Acknowldigements.tex
\begin{credits}
	\subsubsection*{\ackname}
	This publication was produced as part of the ``Zentrum für angewandte Künstliche Intelligenz in Duisburg (ZaKI.D)'' project.
	ZaKI.D is funded under the 5-Standorte Program by the Federal Ministry for Economic Affairs and Energy (BMWi) and the Ministry of Economic Affairs, Industry, Climate Protection, and Energy of the State of North Rhine-Westphalia (MWIKE NRW) (grant number 11-09862).

	\subsubsection*{\discintname}
	The authors have no competing interests to declare that are relevant to the content of this article.
\end{credits}

%% file: references.bib
@inproceedings{bersatti_neural_2021,
  title = {Neural {{Network Weight Compression}} with {{NNW-BDI}}},
  booktitle = {Proceedings of the {{International Symposium}} on {{Memory Systems}}},
  author = {Bersatti, Andrei and Shoghi Ghalehshahi, Nima and Kim, Hyesoon},
  year = 2021,
  month = mar,
  series = {{{MEMSYS}} '20},
  pages = {335--340},
  publisher = {Association for Computing Machinery},
  address = {New York, NY, USA},
  doi = {10.1145/3422575.3422805},
  urldate = {2025-04-28},
  isbn = {978-1-4503-8899-3}
}

@article{buron_reducing_2025,
  title = {Reducing {{Memory}} and {{Computational Cost}} for {{Deep Neural Network Training}} with {{Quantized Parameter Updates}}},
  author = {Buron, Leo and Erbsl{\"o}h, Andreas and Schiele, Gregor},
  year = 2025,
  month = aug,
  journal = {JUCS - Journal of Universal Computer Science},
  volume = {31},
  number = {9},
  pages = {963--979},
  issn = {0948-6968, 0948-695X},
  doi = {10.3897/jucs.164737},
  urldate = {2026-04-24},
  copyright = {http://creativecommons.org/licenses/by/4.0/},
  langid = {english}
}

@inproceedings{einhaus_insitu_2021,
  title = {In-{{Situ Artificial Intelligence}} for {{Self-}}* {{Devices}}: {{The Elastic AI Ecosystem}} ({{Tutorial}})},
  shorttitle = {In-{{Situ Artificial Intelligence}} for {{Self-}}* {{Devices}}},
  booktitle = {2021 {{IEEE International Conference}} on {{Autonomic Computing}} and {{Self-Organizing Systems Companion}} ({{ACSOS-C}})},
  author = {Einhaus, Lukas and Qian, Chao and Ringhofer, Christopher and Schiele, Gregor},
  year = 2021,
  month = sep,
  pages = {320--321},
  doi = {10.1109/ACSOS-C52956.2021.00085},
  urldate = {2026-06-04}
}

@inproceedings{hu_deltadnn_2020,
  title = {Delta-{{DNN}}: {{Efficiently Compressing Deep Neural Networks}} via {{Exploiting Floats Similarity}}},
  shorttitle = {Delta-{{DNN}}},
  booktitle = {Proceedings of the 49th {{International Conference}} on {{Parallel Processing}}},
  author = {Hu, Zhenbo and Zou, Xiangyu and Xia, Wen and Jin, Sian and Tao, Dingwen and Liu, Yang and Zhang, Weizhe and Zhang, Zheng},
  year = 2020,
  month = aug,
  series = {{{ICPP}} '20},
  pages = {1--12},
  publisher = {Association for Computing Machinery},
  address = {New York, NY, USA},
  doi = {10.1145/3404397.3404408},
  urldate = {2025-04-28},
  isbn = {978-1-4503-8816-0}
}

@article{hunt_delta_1998,
  title = {Delta Algorithms},
  author = {Hunt, James J. and Vo, Kiem-Phong and Tichy, W.},
  year = 1998,
  journal = {ACM Transactions on Software Engineering and Methodology (TOSEM)},
  volume = {7},
  pages = {192--214},
  doi = {10.1145/279310.279321}
}

@article{jin_design_2023,
  title = {Design of a {{Quantization-Based DNN Delta Compression Framework}} for {{Model Snapshots}} and {{Federated Learning}}},
  author = {Jin, Haoyu and Wu, Donglei and Zhang, Shuyu and Zou, Xiangyu and Jin, Sian and Tao, Dingwen and Liao, Qing and Xia, Wen},
  year = 2023,
  month = mar,
  journal = {IEEE Transactions on Parallel and Distributed Systems},
  volume = {34},
  number = {3},
  pages = {923--937},
  issn = {1558-2183},
  doi = {10.1109/TPDS.2022.3230840},
  urldate = {2025-04-28}
}

@misc{pekhimenko_practical_2016,
  title = {Practical {{Data Compression}} for {{Modern Memory Hierarchies}}},
  author = {Pekhimenko, Gennady},
  year = 2016,
  month = sep,
  number = {arXiv:1609.02067},
  eprint = {1609.02067},
  primaryclass = {cs.AR},
  publisher = {arXiv},
  doi = {10.48550/arXiv.1609.02067},
  urldate = {2026-05-16},
  archiveprefix = {arXiv}
}

@book{sayood_lossless_2003,
  title = {Lossless Compression Handbook},
  author = {Sayood, {\relax Khalid}.},
  year = 2003,
  series = {Academic {{Press}} Series in Communications, Networking and Multimedia},
  edition = {1st edition},
  publisher = {Academic Press},
  address = {Amsterdam ;},
  isbn = {9786611049881},
  langid = {english}
}

@inproceedings{schiele_elastic_2019,
  title = {The {{Elastic Node}}: {{An Experimentation Platform}} for {{Hardware Accelerator Research}} in the {{Internet}} of {{Things}}},
  booktitle = {2019 {{IEEE International Conference}} on {{Autonomic Computing}} ({{ICAC}})},
  author = {Schiele, Gregor and Burger, Alwyn and Cichiwskyj, Christopher},
  year = 2019,
  month = jun,
  pages = {84--94},
  issn = {2474-0756},
  doi = {10.1109/ICAC.2019.00020}
}

@misc{xiao_fashionmnist_2017,
  title = {Fashion-{{MNIST}}: A {{Novel Image Dataset}} for {{Benchmarking Machine Learning Algorithms}}},
  shorttitle = {Fashion-{{MNIST}}},
  author = {Xiao, Han and Rasul, Kashif and Vollgraf, Roland},
  year = 2017,
  month = sep,
  number = {arXiv:1708.07747},
  eprint = {1708.07747},
  primaryclass = {cs},
  doi = {10.48550/arXiv.1708.07747},
  urldate = {2026-05-06},
  archiveprefix = {arXiv}
}

@inproceedings{zhang_qdcompressor_2021,
  title = {{{QD-Compressor}}: A {{Quantization-based Delta Compression Framework}} for {{Deep Neural Networks}}},
  shorttitle = {{{QD-Compressor}}},
  booktitle = {2021 {{IEEE}} 39th {{International Conference}} on {{Computer Design}} ({{ICCD}})},
  author = {Zhang, Shuyu and Wu, Donglei and Jin, Haoyu and Zou, Xiangyu and Xia, Wen and Huang, Xiaojia},
  year = 2021,
  month = oct,
  pages = {542--550},
  issn = {2576-6996},
  doi = {10.1109/ICCD53106.2021.00088},
  urldate = {2025-04-28}
}
